\documentclass[a4paper, 11pt]{amsart}
\usepackage{amsfonts}

\begin{document}

\title{On the Tomographic Picture of Quantum Mechanics}

\author{A. Ibort$^1$, V.I. Man'ko$^2$, G. Marmo$^3$, A. Simoni$^3$, F. Ventriglia$^3$}

%\thanks{This work was partially supported by MTM2007-62478 Research project, Ministry of Science, Spain.}

%\date{\today}

\maketitle

{\parindent 0cm {\footnotesize $^1$\textit{Departamento de Matem\'{a}ticas,
Universidad Carlos III de Madrid, Avda. de la Universidad 30, 28911
Legan\'{e}s, Madrid, Spain.}  (\texttt{albertoi@math.uc3m.es})}

{\footnotesize \textit{$^2$P.N.Lebedev Physical Institute,
Leninskii Prospect 53, Moscow 119991, Russia.} (\texttt{manko@na.infn.it})}

{\footnotesize \textit{$^3$Dipartimento di Scienze Fisiche dell' Universit\`{a} ``Federico II" e Sezione INFN di Napoli, Complesso Universitario di Monte S. Angelo, via Cintia, 80126 Naples, Italy.}  (\texttt{marmo@na.infn.it,
simoni@na.infn.it, ventriglia@na.infn.it})}
}

\begin{abstract}
We formulate necessary and sufficient conditions for a symplectic
tomogram of a quantum state to determine the density state. We
establish a connection between the (re)construction by means of
symplectic tomograms  with the construction by means of Naimark
positive-definite functions on the Weyl-Heisenberg group. This
connection is used to formulate properties which guarantee that
tomographic probabilities describe quantum states in the
probability representation of quantum mechanics.
\end{abstract}



\markboth{Positive quantum tomograms}{Positive quantum tomograms}

\newpage
%%%%%%%%%%%%%%%%%%%%%%%%%%%%%%%%%%%%%%%%%%%%%%%%

\markboth{Positive quantum tomograms}{Positive quantum tomograms}

%%%%%%%%%%%%%%%%%%%%%%%%%%%%%%%%%%%%%%%%%%%%%%%%%%%%%%%%%%%%%%%%%%%%%%%%%%%%%%%%%
\section{Introduction}

It has been shown recently (\cite{Mancini,Found Phys
97,Ventriglia2}, see also \cite{Pedatom}) how to describe quantum
states by using a standard positive probability distribution
called a symplectic probability distribution or symplectic
tomogram. The symplectic tomogram $\mathcal{W}(X,\mu ,\nu )$ is a
nonnegative function of the random position $X$ measured in
reference frames
in phase--space with rotated and scaled axes $q\rightarrow \mu q$, $%
p\rightarrow \nu p$ where $\mu =e^{\lambda }\cos \theta$, $\nu
=e^{-\lambda }\sin \theta $, $\theta $ is the angle of rotation
and $e^{\lambda }$ is the scaling parameter.

The symplectic tomographic probability distribution
$\mathcal{W}(X,\mu ,\nu ) $ contains complete information on
quantum states in the sense that for a given wave function $\psi
(x)$ or density operator $\hat{\rho}$ (determining the quantum
state \cite{Beppe_book}, \cite{VonNeumann27} \ in the conventional
formulation of quantum mechanics) the tomogram can be calculated.
On the other hand for a given tomogram $\mathcal{W}(X,\mu ,\nu )$
one can reconstruct explicitly the density operator $\hat{\rho}$.
It means that for a given symplectic tomogram of a system with
continuous variables all the properties of the quantum system can
be obtained as well as for a given density operator $\hat{\rho}$.
Analogous complete information on the quantum states is contained
in the Wigner function \cite{Wig32} $W(q,p)$ which is a real
function on the phase space of the system. The Wigner function is
related to the symplectic tomogram by means of an integral Radon
transform \cite{Radon1917}, however the Wigner function is not
definite in sign, it takes negative values for some quantum states
and cannot be considered as a positive probability distribution on
phase space. The necessary and sufficient conditions for a real
function on the phase space to describe the Wigner function of a
quantum state were found in \cite{Nar86} where the corresponding
properties of the function under consideration were associated
with the so called $h$--positivity condition of a function on the
Abelian translation group on the phase space.

As we have shown elsewhere \cite{Ib09}, in this description plays
an important role the Weyl-Heisenberg group and its group of
automorphisms, along with the Abelian vector group which arises as
quotient group of Weyl-Heisenberg group by its central subgroup.
In this paper we would like to consider the tomographic
description of quantum mechanics as another picture, on the same
footing as the Schroedinger, Heisenberg or Weyl--Wigner pictures.
To this aim we have to provide a characterization of symplectic
tomograms which stands on its own, without relying on other
pictures. In other terms, we need necessary and sufficient
conditions for a function $f(X,\mu ,\nu )$ to be the symplectic
tomogram  $\mathcal{W}(X,\mu ,\nu )$ of a quantum state. The
strategy to find these conditions is based on Naimark's theorem
\cite{Naim Book} that provides a characterization of positive
operator valued measures and that allows to characterize functions
which are elements of matrices of
group representations. In particular we use the result that a function $%
\varphi (g)$ on a group $G$, $g\in G$, which is a diagonal matrix
element of a unitary representation of the group $G$ has the
property of being positive definite in the sense that the matrix
\begin{equation}
M_{jk}=\varphi (g_{j}g_{k}^{-1})
\end{equation}
for any $j,k=1,2,...,N$ and arbitrary $N$, is positive definite.
Below we will show that symplectic tomograms can be associated
with positive definite functions $\varphi $ on the Weyl-Heisenberg
group. Since Naimark's theorem for positive operator--valued
measures allows to construct and determine
uniquely a Hilbert space and a vector in it representing the function $%
\varphi $ (using what today is called the Gelfand-Naimark-Segal
(GNS) method) the connection established below of the symplectic
tomograms with positive definite functions on the Weyl-Heisenberg
group yields the necessary and sufficient condition which we are
looking for.

It is worthy to note that this condition can be also studied using
the necessary and sufficient condition for a function to be a
Wigner function \cite{Nar86} but we do not use here the connection
of symplectic tomogram with the Wigner function and provide the
condition for the tomogram independently of any other result
concerning Wigner functions.

\bigskip

\section{Symplectic tomography}

In this section we briefly review the construction of tomographic
probability densities determining the quantum state of a particle
in one degree of freedom \cite{Pedatom}. Given the density
operator $\hat{\rho}$ of
a particle quantum state, $\hat{\rho}=\hat{\rho}^{\dagger }$, $\mathrm{Tr}%
\hat{\rho}=1$, and $\hat{\rho}\geq 0$, the symplectic tomogram of
$\hat{\rho} $ is defined by:
\begin{equation}
\mathcal{W}(X,\mu ,\nu )=\mathrm{Tr}[\hat{\rho}\,\delta (X\,\hat{1}-\mu \hat{%
Q}-\nu \hat{P})],\qquad X,\mu ,\nu \in \mathbb{R}.  \label{def
tom}
\end{equation}
Here $\hat{Q}$ and $\hat{P}$ are the position and momentum
operators. The Dirac delta--function with operator arguments is
defined by the standard Fourier integral,
\begin{equation*}
\delta (X\,\hat{1}-\mu \hat{Q}-\nu \hat{P})=\int e^{-ik(X\,\hat{1}-\mu \hat{Q%
}-\nu \hat{P})}\frac{dk}{2\pi }.
\end{equation*}
The symplectic tomogram $\mathcal{W}(X,\mu ,\nu )$ has the
properties which follow from its definition by using the known
properties of delta-function, namely:

\begin{enumerate}
\item[i.]  Nonnegativity:
\begin{equation}
\mathcal{W}(X,\mu ,\nu )\geq 0  \label{tom_pos}
\end{equation}
(this holds by observing that the trace of the product of two
positive operators is a positive number).

\item[ii.]  Normalization:
\begin{equation}
\int \mathcal{W}(X,\mu ,\nu )dX=1.  \label{tom_normalization}
\end{equation}

\item[iii.]  Homogeneity:
\begin{equation}
\mathcal{W}(\lambda X,\lambda \mu ,\lambda \nu )=\frac{1}{\left|
\lambda \right| }\mathcal{W}(X,\mu ,\nu ).
\label{tom_homogeneity}
\end{equation}
\end{enumerate}

However, the three above properties are by no means sufficient to
determine the quantum character of a tomographic function $f(X,\mu
,\nu ).$ For instance, consider
\begin{equation}
f(X,\mu ,\nu )=\exp \left( -\frac{X^{2}}{2\left( \mu ^{2}+\nu ^{2}\right) }%
\right) \frac{5\left( \mu ^{2}+\nu ^{2}\right)
-X^{2}}{\sqrt{2\left( \mu ^{2}+\nu ^{2}\right) ^{3}}}.
\end{equation}
Despite the uncertainty relations are satisfied by such a
function, $f$ is not a quantum tomogram because $\left\langle
P^{2}\right\rangle =\left\langle Q^{2}\right\rangle =-1/2$, as it
can be checked using
\begin{equation}
\left\langle P^{2}\right\rangle =\int X^{2}\left. f(X,\mu ,\nu
)\right| _{\mu =0,\nu =1}dX
\end{equation}
and analogously for $\left\langle Q^{2}\right\rangle .$

On the other hand, it is easy to see that formula (\ref{def tom})
has an inverse \cite{QSO1996}:
\begin{equation}
\hat{\rho}=\frac{1}{2\pi }\int \mathcal{W}(X,\mu ,\nu
)e^{i(X\,\hat{1}-\mu \hat{Q}-\nu \hat{P})}\,dX\,d\mu \,d\nu .
\label{tom_inversion}
\end{equation}
Thus the knowledge of the symplectic tomogram $\mathcal{W}(X,\mu
,\nu )$ means that the density operator $\hat{\rho}$ is also
known, more precisely,
can be reconstructed. This correspondence between symplectic tomograms $%
\mathcal{W}(X,\mu ,\nu )$ and density operators $\hat{\rho}$ gives
the possibility to formulate the notion of quantum state using
tomograms as the primary notion. However to make this idea
precise, we need to formulate additional conditions to be
satisfied by the function $\mathcal{W}(X,\mu ,\nu )$ which are
extra to the conditions (\ref{tom_pos})-(\ref{tom_homogeneity})
and which guarantee that by using the inversion formula
(\ref{tom_inversion}) we get an operator with all the necessary
properties of a density state. The general recipe to formulate
these demands can be given by checking the nonnegativity condition
of the integral (see \cite{Pedatom}):
\begin{equation}
\int \mathcal{W}(X,\mu ,\nu )e^{i(X\,\hat{1}-\mu \hat{Q}-\nu \hat{P}%
)}\,dX\,d\mu \,d\nu \geq 0.  \label{inverse_op}
\end{equation}
It means that for a given function $\mathcal{W}(X,\mu ,\nu )$
satisfying the conditions (\ref{tom_pos})-(\ref{tom_homogeneity})
one has to check the
nonnegativity of the operator (\ref{inverse_op}), thus if the inequality (%
\ref{inverse_op}) holds the function $\mathcal{W}(X,\mu ,\nu )$ is
the symplectic tomogram of a quantum state, however it must be
realized that this is not an operative procedure.

Below we formulate the conditions for a function
$\mathcal{W}(X,\mu ,\nu )$ to be a tomogram of a quantum state
avoiding the integrations in eq. (\ref {inverse_op}). As
anticipated in the introduction, to be able to use Naimark's
results we have to deal with functions defined on a group. Thus,
we have to show how symplectic tomograms may be associated with
the Weyl-Heisenberg group. In doing this we can exploit results in
\cite{Naim Book} where the theorems on properties of diagonal
matrix elements of unitary representations provide the key to
construct tomograms which represent quantum states.

\section{Tomographic probability measures}

To get a  mathematical formulation of the tomographic picture we
invoke the spectral theory of Hermitian operators, which moreover
will provide us with a probabilistic interpretation of the
symplectic tomogram. We start rewriting the formal definition, eq.
(\ref{def tom}), for a quantum tomogram:
\begin{equation}
\mathcal{W}(X,\mu ,\nu )=\mathrm{Tr}\left[ \hat{\rho}\int e^{ik(X\,\hat{1}%
-\mu \hat{Q}-\nu \hat{P})}\frac{dk}{2\pi }\right] =\int e^{ikX}\mathrm{Tr}[%
\hat{\rho}e^{-ik(\mu \hat{Q}+\nu \hat{P})}]\frac{dk}{2\pi }.
\label{def_tomo}
\end{equation}
then we observe that
\begin{equation}
\mu \hat{Q}+\nu \hat{P}=S_{\mu \nu }\hat{Q}S_{\mu \nu }^{\dagger }
\end{equation}
where
\begin{equation}
S_{\mu \nu }=\exp \left[ \frac{i\lambda }{2}\left( \hat{Q}\hat{P}+\hat{P}%
\hat{Q}\right) \right] \exp \left[ \frac{i\theta }{2}\left( \hat{Q}^{2}+\hat{%
P}^{2}\right) \right] ,
\end{equation}
with
\begin{equation}
\mu =e^{\lambda }\cos \theta \ ,\nu =e^{-\lambda }\sin \theta \ .
\end{equation}
In other words, by acting with the unitary operators $S_{\mu \nu
}$ on the position operator $\hat{Q}$ we get out the iso-spectral
family of hermitian operators
\begin{equation*}
X_{\mu \nu }=\mu \hat{Q}+\nu \hat{P}.
\end{equation*}
This family is a symplectic tomographic set \cite{CSTom}.

To any operator of this family is associated a projector valued
measure $\Pi _{\mu \nu }$ on the $\sigma $--algebra of Borel sets
on the real line:
\begin{equation*}
\mu \hat{Q}+\nu \hat{P}=\int \lambda \,d\,\Pi _{\mu \nu }(\lambda
).
\end{equation*}
Given any density state $\hat{\rho}$, the projector valued measure
$\Pi _{\mu \nu }$ yields a normalized probability measure $m_{\rho
,\mu \nu }$ on the Borel sets $E\in \mathrm{Bo}(\mathbb{R})$ of
the real line:
\begin{equation}
m_{\rho ,\mu \nu }(E)=\mathrm{Tr}[\,\hat{\rho}\,\Pi _{\mu \nu
}(E)];\quad m_{\rho ,\mu \nu }(\mathbb{R})=1.
\end{equation}
We recall that $m_{\rho ,\mu \nu }(E)$ is the probability that a
measure of
the observable $\mu \hat{Q}+\nu \hat{P}$ in the state $\hat{\rho}$ is in $E$%
. All these measures $m_{\rho ,\mu \nu }$ are absolutely
continuous with
respect to the Lebesgue measure $dX$ on the real line, so that densities $%
V_{\rho }(X,\mu ,\nu )$ may be introduced such that
\begin{equation}
m_{\rho ,\mu \nu }(E)=\int_{E}V_{\rho }(X,\mu ,\nu )\,dX.
\end{equation}
We can write
\begin{equation}
\mathrm{Tr}\left( \hat{\rho}e^{-i\lambda (\mu \hat{Q}+\nu \hat{P})}\right) =%
\mathrm{Tr}\left( \hat{\rho}S_{\mu \nu }e^{-i\lambda
\hat{Q}}S_{\mu \nu }^{\dagger }\right) =\int e^{-i\lambda X\
}V_{\rho }(X,\mu ,\nu )\,dX \label{fun caratt}
\end{equation}
so that
\begin{eqnarray}
\mathcal{W}(X,\mu ,\nu ) &=&\int e^{ikX\
}\mathrm{Tr}[\hat{\rho}e^{-ik(\mu
\hat{Q}+\nu \hat{P})}]\frac{dk}{2\pi } \\
&=&\int e^{ikX\ }e^{-ikX\ ^{\prime }}V_{\rho }(X^{\prime },\mu
,\nu
)dX^{\prime }\frac{dk}{2\pi }\notag \\
&=&\int \delta (X-X^{\prime })V_{\rho }(X^{\prime },\mu ,\nu
)dX^{\prime }=V_{\rho }(X,\mu ,\nu ).  \notag
\end{eqnarray}
In other words we have shown that the symplectic tomogram
$\mathcal{W}(X,\mu ,\nu )$ of a given state $\hat{\rho}$ is
nothing but the density $V_{\rho }(X,\mu ,\nu )$ of the
probability measure associated to the state by means of the
symplectic tomographic set. The tomographic character of the
family of observables $X_{\mu \nu }$ is contained in the
possibility of reconstructing any state out of the corresponding
probability measures by means of the previous reconstructing
formula. By using eqs. (\ref{fun caratt}) and
(\ref{tom_inversion}), we get
\begin{equation}\label{lineremoved}
\hat{\rho} =
=\frac{1}{2\pi }\int \mathrm{Tr}[\hat{\rho}e^{i(\mu \hat{Q}+\nu \hat{P}%
)}]\,e^{-i(\mu \hat{Q}+\nu \hat{P})}\,d\mu \,d\nu ,
\end{equation}
moreover
\begin{equation}
\frac{1}{2\pi }\int \mathrm{Tr}[\,e^{i(\mu \hat{Q}+\nu
\hat{P})}]\,e^{-i(\mu \hat{Q}+\nu \hat{P})}\,d\mu \,d\nu =
\hat{1}.
\end{equation}
The presence of Weyl operators $D(\mu ,\nu )=e^{i(\mu \hat{Q}+\nu
\hat{P})}$ suggest that we are dealing with projective
representations of the Abelian vector group. We shall take up
group theoretical aspects in next section.

\section{A group theoretical description of quantum tomograms}

The probabilistic interpretation above allows to consider the
tomographic description of quantum states as a picture of quantum
mechanics on the same footing as other well known representations,
like Schr\"{o}dinger, Heisenberg and Wigner-Weyl for instance.
Thus, to be an alternative picture of quantum mechanics we need
criteria to recognize a function $f(X,\mu ,\nu ) $ as a tomogram
of a quantum state. For this, the use of the reconstruction
formula to check if the obtained operator is a density operator
would be unsatisfactory, mainly because this check requires to
switch from tomographic to Schr\"{o}dinger picture. In other
words, we would like to establish self-contained criteria for a
function to be a quantum tomogram. More precisely, we have to
address the following problem: given a tomogram-like function
$f(X,\mu ,\nu ),$ that is a function with the above properties
eqs. (\ref{tom_pos})-(\ref{tom_homogeneity}) of a tomogram, what
are the necessary and sufficient  conditions to recognize $f$ as a
quantum tomogram?

To this aim we begin to observe that in the characteristic
tomographic function
\begin{equation}
\mathrm{Tr}[\hat{\rho}e^{i(\mu \hat{Q}+\nu \hat{P})}]=\mathrm{Tr}[\hat{\rho}%
D(\mu ,\nu )]
\end{equation}
a projective representation of the translation group appears. This
projective representation can be lifted to a true unitary
representation of the Weyl-Heisenberg group (see for instance
\cite{Ib09} and references therein for a detailed discussion of
the subject) by means of a central extension of the translation
group. Such central extension defines the Weyl-Heisenberg group
\textit{WH}$(2)$ whose elements are denoted by $(\mu ,\nu ,t)$ and
the group law reads:
\begin{equation}
(\mu ,\nu ,t)\circ (\mu ^{\prime },\nu ^{\prime },t^{\prime
})=(\mu +\mu ^{\prime },\nu +\nu ^{\prime },t+t^{\prime
}+\frac{1}{2}\omega ((\mu ,\nu ),(\mu ^{\prime },\nu ^{\prime
}))),
\end{equation}
where $\omega ((\mu ,\nu ),(\mu ^{\prime },\nu ^{\prime }))=\mu
\nu ^{\prime }-\nu \mu ^{\prime }$ denotes the symplectic form on
$\mathbb{R}^{2}$. The nontrivial unitary irreducible
representations of the Weyl-Heisenberg group are provided by the
expression:
\begin{equation}
U_{\gamma }(\mu ,\nu ,t)=D_{\gamma }(\mu ,\nu )e^{i\gamma tI}.
\label{reppol}
\end{equation}
where $\gamma $ is a non-vanishing real number.In what follows we
will set $\gamma=1 $. Hence we immediately observe that
\begin{equation}
\mathrm{Tr}[\hat{\rho}D(\mu ,\nu )]=e^{- it}\mathrm{Tr}[\hat{\rho}%
U(\mu ,\nu ,t)]  \label{tra_uni}
\end{equation}
where the function $\mathrm{Tr}[\hat{\rho}U(\mu ,\nu ,t)]$ is of
positive type \cite{Naim Book}.

For convenience we recall the definition of functions of positive
type. Given a group $G$ a function $\varphi (g)$ on $G$ ($g\in G$
) is of positive type,
or definite positive, if for any $n$--tuple of group elements $%
(g_{1},g_{2},...,g_{n})$ the matrix
\begin{equation}
M_{jk}=\varphi (g_{j}g_{k}^{-1})\quad \quad j,k=1,2,...,n,
\label{Nay matrix}
\end{equation}
is positive semi-definite for any $n\in \mathbb{N}$, or in other
words, if for any finite family of elements $g_{1},g_{2},\ldots
,g_{n}\in G$ and for
any family of complex numbers $\xi _{1},\ldots ,\xi _{n}$, we have $%
\sum_{j,k=1}^{n}\bar{\xi _{j}}\xi _{k}\varphi
(g_{j}g_{k}^{-1})\geq 0$, for any $n$. Moreover, a simple
computation shows that given any unitary
representation $U(g)$ of $G$ and a state $\rho $, $\mathrm{Tr}[\hat{\rho}%
U(g)]$ \ is a group function of positive type. \textsl{Viceversa
}any positive type group function $\varphi (g)$ can be written in
the form
\begin{equation}
\mathrm{Tr}[\hat{\rho}_{\xi }U(g)]=\langle \xi ,U(g)\xi \rangle ,
\label{naimark}
\end{equation}
where $U(g)$ is a unitary representation and $\xi $ is a cyclic
vector in a suitable Hilbert space, obtained for instance by means
of a GNS construction
\cite{Naim Book}. Thus the condition on the matrix introduced in (\ref{Nay matrix}) is a way to affirm that $%
\varphi $ is associated with a state without making recourse to a
representation.

Thus we can state the required condition: a tomogram--like
function $f(X,\mu ,\nu )$ is a quantum tomogram, i.e., there
exists a quantum state $\hat{\rho} $ such that $f(X,\mu ,\nu
)=\mathrm{Tr}[\hat{\rho}\,\delta (X\,\hat{1}-\mu \hat{Q}-\nu
\hat{P})]$, if and only if its Fourier transform evaluated at 1
may be written in the form
\begin{equation}
\int f(X,\mu ,\nu )e^{iX}dX=e^{-it}\varphi _{f}(\mu ,\nu ,t)
\label{Fourier group}
\end{equation}
where $\varphi _{f}(\mu ,\nu ,t)$ is a positive definite function
on the Weyl-Heisenberg group. In fact if $\ \mathcal{W}$ is a
quantum tomogram, then because of eqs. (\ref{def_tomo}) and
(\ref{tra_uni}) we have,
\begin{equation}
\int \mathcal{W}(X,\mu ,\nu
)e^{iX}\,dX=\mathrm{Tr}[\hat{\rho}D(\mu ,\nu
)]=e^{-it}\mathrm{Tr}[\hat{\rho}U(\mu ,\nu ,t)]=e^{-it}\varphi
(\mu ,\nu ,t),
\end{equation}
where $\varphi (\mu ,\nu ,t)$ is a positive definite function on
the
Weyl-Heisenberg group. Moreover if we define $\psi (\mu ,\nu )=\mathrm{Tr%
}(\hat{\rho}D(\mu ,\nu )$, then $\psi (\mu ,\nu )$ is a function
on the translation group considered as a quotient of the
Weyl-Heisenberg group by the central element. It means that we are
dealing with a projective representation and not a unitary
representation like in Naimark's theorem eq. (\ref{naimark}). Then
 we could ask about the properties enjoyed by the matrix
$\tilde{M}_{jk}$
constructed using $\psi $ instead of $\varphi $. If we denote as above by $%
\omega $ the 2--cocycle defining the projective representation,
then we will say that $\tilde{M}_{jk}$ is of \ $\omega $--positive
type, i.e.
\begin{equation}
\tilde{M}_{jk}=\psi ((\mu _{j},\nu _{j})^{-1}\circ (\mu _{k},\nu _{k}))e^{i%
\frac{1}{2}\omega ((\mu _{k},\nu _{k}),(\mu _{j},\nu _{j})}
\end{equation}
is positive semidefinite. This yields the corresponding condition:
a tomogram--like function $f(X.\mu ,\nu )$ is a quantum tomogram
if and only if its Fourier transform evaluated at 1 may be written
in the form
\begin{equation}
\int f(X,\mu ,\nu )e^{iX}dX=\psi _{f}(\mu ,\nu )
\end{equation}
where $\psi _{f}(\mu ,\nu )$ is a function of the translation group of $%
\omega -$positive type.

We observe that $\psi _{f}(\mu ,\nu )$ may be at same time of positive and $%
\omega -$positive type on the translation group. Then by Bochner theorem $%
\psi _{f}(\mu ,\nu )$ is the Fourier transform of a probability
measure on the phase space. In other words $f(X,\mu ,\nu )$ is the
(classical) Radon transform of such a probability measure i.e. a
classical tomogram. The tomogram of the ground state of the
harmonic oscillator provides an example of the above situation. In
that case, the GNS construction yields a Hilbert space of square
integrable functions on phase space  with respect to the measure
provided by the Bochner theorem.

To finish this analysis let us notice that if $\psi $ is a function of $%
\omega $--positive type on the translation group, then the
function $\varphi (\mu ,\nu ,t)=e^{it}\psi (\mu ,\nu )$ will be a
positive definite function on the Weyl-Heisenberg group
$\mathcal{WH}(2)$ and, by Naimark's theorem, there will exist a
unitary representation $U$ of $\mathcal{WH}(2)$ and a cyclic state
vector $|\xi \rangle $ such that $\varphi (\mu ,\nu ,t)=\langle
\xi ,U(\mu ,\nu ,t)\xi \rangle $ . On the other hand, $\psi (\mu
,\nu )$ is obtained by $f(X,\mu ,\nu )$, which is a tomogram of a
quantum state $\hat\rho .$ Up to a unitary transformation
$\hat\rho $ will coincide with $\hat\rho _{\xi }$ iff it is a pure
state. Notably, the purity of $\hat\rho $ can be expressed as:
\begin{equation}
\mathrm{tr}\hat\rho ^{2}=\frac{1}{2\pi }\int \mathcal{W}(X,v)%
\mathcal{W}(Y,-v)e^{i(X+Y)}dXdYdv=\frac{1}{2\pi}\int_{%
\mathbb{R}^{2}}\left| \psi (v)\right| ^{2}dv
\end{equation}
so that the above condition can be stated as:
\begin{equation}
\int_{\mathbb{R}^{2}}\left| \psi (v)\right| ^{2}dv=1
\label{purity}
\end{equation}
Here vector $v(\mu,\nu)$. The case of a mixed density state $\rho
,$ when eq.(\ref{purity}) does not hold, will be discussed
elsewhere.

\section{Conclusions and outlooks}

To conclude we resume the main results of our work. The symplectic
tomographic probability distribution considered as the primary
concept of a particle quantum state alternative to the wave
function or density matrix, we have shown to be associated with a
unitary representation of the Weyl-Heisenberg group. This
connection was used to formulate the autonomous conditions for the
symplectic tomogram to describe quantum states using the
positivity properties of the matrix $M_{jk}$ of eq.(\ref{Nay
matrix}) introduced in \cite{Naim Book} and connected with
diagonal elements of the unitary representation (positive-type
function $\varphi (g)$ on the group). The function $f(X,\mu ,\nu
),$ satisfying the necessary properties of tomographic probability
distribution, i.e. non-negativity, homogeneity and normalization,
was shown to be a quantum tomogram iff its Fourier transform in
the quadrature variable $X$ can be written in the form of eq.(\ref
{Fourier group}) as the product of a positive type function on the
Weyl-Heisenberg group and a phase factor associated with a central
element of the group. By using the quantum Radon anti-transform
eq.(\ref {tom_inversion}), this condition guarantees that the
function $f(X,\mu ,\nu ) $ provides a density state, so that $f$
is the symplectic tomogram of a quantum state. The criterion,
formulated in terms of positivity properties of a group function
obtained from the tomographic function, is not easy to implement
operatively.Nevertheless, it is simpler than the criterion based
on checking the non negativity of the operator given by the
quantum Radon anti transform. Also we have shown that the purity
of the quantum state can be expressed as the square of the
$L_{2}-$ norm of that positive group function, which is obtained
by tomograms measured directly in optical experiments, without
considering density matrices or Wigner functions. As a spin-off we
have shown that the notion of $h-$ positivity may be subsumed
under the notion of positivity for a centrally extended group. In
this paper we have considered tomograms associated with the
Weyl-Heisenberg group. In a forthcoming paper we will show how to
deal with the tomographic  picture for general Lie groups and for
finite groups. In this connection we shall also elaborate more on
the $C^{\ast }-$algebraic approach to quantum mechanics and its
counterpart in terms of tomograms.

\section*{Acknowledgements}
This work was partially supported by MTM2007-62478 Research
project, Ministry of Science, Spain.
%%%%%%%%%%%%%%%%%%%%%%%%%%%%%%%%%%%%%%%%%%%%%%%%%%%%%%%%%%%%%%%%%%%%%%%%%%%%%%%%%%%%%%%%%%%%%%%%

\end{document}